\pdfoutput=1

\documentclass[11pt]{article}

\usepackage[preprint]{acl}
\usepackage{times}
\usepackage{latexsym}

\usepackage[T1]{fontenc}

\usepackage[utf8]{inputenc}

\usepackage{microtype}

\usepackage{inconsolata}
\usepackage{algorithm}
\usepackage{algorithmic}
\usepackage{amsmath}
\usepackage{stfloats}
\usepackage{caption}
\usepackage{tabularx}
\usepackage{stfloats}
\usepackage{amssymb}
\usepackage[figuresright]{rotating}
\usepackage{lineno}
\usepackage{hyperref}
\usepackage{bm}
\usepackage{subfigure}
\usepackage{multirow}
\usepackage{pifont}
\usepackage{mathrsfs}
\usepackage{makecell}
\usepackage{color}
\usepackage{array}
\usepackage{graphicx}
\usepackage{rotating}

\title{Enhancing Emotion Recognition in Incomplete Data: A Novel Cross-Modal Alignment, Reconstruction, and Refinement Framework}

\author{\quad Haoqin Sun$^{1}$ \quad Shiwan Zhao$^{1}$ \quad Shaokai Li$^{2}$ \quad  Xiangyu Kong$^{3}$ \quad  Xuechen Wang$^{1}$ \\
\textbf{\quad Aobo Kong$^{1}$ \quad Jiaming Zhou$^{1}$ \quad Yong Chen$^{4}$ \quad Wenjia Zeng$^{4}$ \quad Yong Qin$^{1}$}\thanks{~~Yong Qin is the corresponding author.}\\
$^1$Nankai University \quad $^2$Independent Researcher\\
$^3$University of Leicester \quad $^4$Lingxi (Beijing) Technology Co., Ltd.\\
\texttt{$^{1}$sunhaoqin@mail.nankai.edu.cn} }

\begin{document}
\maketitle
\begin{abstract}
Multimodal emotion recognition systems rely heavily on the full availability of modalities, suffering significant performance declines when modal data is incomplete. To tackle this issue, we present the Cross-Modal Alignment, Reconstruction, and Refinement (CM-ARR) framework, an innovative approach that sequentially engages in cross-modal alignment, reconstruction, and refinement phases to handle missing modalities and enhance emotion recognition. This framework utilizes unsupervised distribution-based contrastive learning to align heterogeneous modal distributions, reducing discrepancies and modeling semantic uncertainty effectively. The reconstruction phase applies normalizing flow models to transform these aligned distributions and recover missing modalities. The refinement phase employs supervised point-based contrastive learning to disrupt semantic correlations and accentuate emotional traits, thereby enriching the affective content of the reconstructed representations. Extensive experiments on the IEMOCAP and MSP-IMPROV datasets confirm the superior performance of CM-ARR under conditions of both missing and complete modalities. Notably, averaged across six scenarios of missing modalities, CM-ARR achieves absolute improvements of 2.11\% in WAR and 2.12\% in UAR on the IEMOCAP dataset, and 1.71\% and 1.96\% in WAR and UAR, respectively, on the MSP-IMPROV dataset. 
\end{abstract}

\section{Introduction}

Multimodal emotion recognition (MMER) entails the analysis of emotional cues across various modalities, including speech, text, and body language, among others. These modalities serve complementary functions in the expression and interpretation of human emotions. However, in practical applications, the availability of these modalities is frequently compromised; specific modalities may be absent or inaccessible due to various factors. For example, text data may be unavailable due to errors in automatic speech recognition systems, speech may be obscured by excessive background noise, and visual data may be impaired by poor lighting or occlusions. These challenges underscore the need for MMER systems to be highly adaptable and robust, capable of effectively functioning even with incomplete modal information.

\begin{figure}[t]
  \centering
  \includegraphics[width=3.0in]{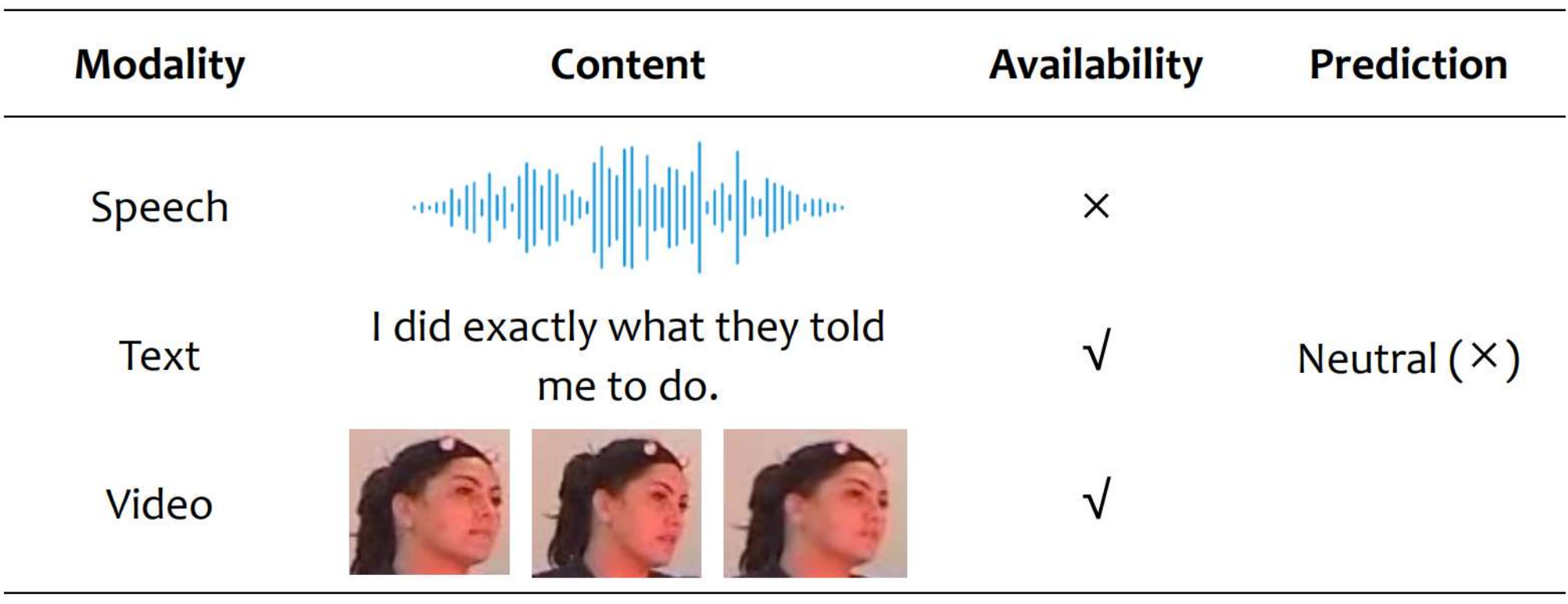}
  \caption{An example of missing modalities: when the speech modality is missing, emotion recognition is guided by the text and video modalities, leading to incorrect predictions. The ground truth is \emph{``angry''}.}
  \label{fig:framework1}
\end{figure}

Conventional multimodal learning paradigms, as documented in the literature \cite{yoon2018multimodal,liu2022multi,sun2024fine}, typically operate under the assumption of complete modality presence. These approaches are dedicated to constructing fusion models optimized for scenarios where all modalities are fully available, a presumption that can undermine their utility in situations where modalities are partially missing. Illustratively, as depicted in Fig.~\ref{fig:framework1}, an emotion classified as \emph{``angry''} in a fully modal context—primarily due to pronounced tones in the speech modality—may be reinterpreted as \emph{``neutral''} in the absence of the speech component, with the text and video modalities becoming the guide.

The research community has developed innovative methodologies aimed at enhancing the resilience of MMER systems faced with incomplete modalities, primarily focusing on predicting missing data across modalities. However, the significant distribution gaps between different modalities present substantial challenges. For instance, \citet{wang2023distribution} utilize flow models to map heterogeneous modality representations into a latent space following a Gaussian distribution, aiming to ensure consistency. Nevertheless, this approach does not adequately address the distribution gaps between modalities. Alternatively, \citet{zuo2023exploiting} suggest the use of modality-invariant features to aid in reconstructing modality-specific characteristics. While this method is promising, it has been critiqued for its limited effectiveness, particularly due to difficulties in accurately predicting modality-specific features. These strategies underscore the critical need for effective cross-modal alignment before attempting to predict missing data across modalities.

To address the aforementioned problems, this paper introduces a novel framework for cross-modal alignment, reconstruction, and refinement, designated as \textbf{CM-ARR}. The initial alignment phase aims to bridge the distributional divergences between modalities, which facilitates subsequent reconstruction efforts. Specifically, inspired by MAP \cite{ji2023map}, we employ an \textbf{unsupervised\footnote{Modalities from the same instance are treated as positive samples without relying on explicit class labels.} distribution-based contrastive learning} approach that replaces point representations with their Gaussian distributional counterparts. This method can convey richer multimodal semantic information by effectively encoding uncertainty. Next in the reconstruction phase, we deploy a network based on normalizing flow models. This method transforms aligned modality representations into a Gaussian latent space \cite{wang2023distribution}. Gaussian distributions for missing modalities are estimated by transferring characteristics from available modalities. Subsequently, the representation of the missing modality is obtained through inverse normalization of the estimated Gaussian distribution.
In the final refinement phase, we refine modality representations to better capture emotional characteristics. The initial phases prioritize semantic alignment, sidelining emotional attributes. Consequently, we implement \textbf{supervised point-based contrastive learning}. This method considers modalities from different instances of the same class as positive samples and thus disrupts the semantic correlation between modalities. Doing so enables the model to capture emotional attributes and related features beyond mere semantics, enhancing the reconstruction of emotional information.

Overall, CM-ARR begins by aligning modalities to harmonize disparate modal distributions, a step that facilitates the effective estimation of missing data in the subsequent reconstruction phase, and concludes with refinement to accentuate emotional traits. The key contributions of this paper are summarized as follows:
\begin{itemize}
\item We introduce the CM-ARR framework, a pioneering approach for cross-modal alignment, reconstruction, and refinement, designed to enhance emotion recognition in scenarios characterized by incomplete data.
\item We present two contrastive learning strategies: unsupervised distribution-based contrastive learning for effective uncertainty modeling and mitigation of distributional disparities, alongside supervised point-based contrastive learning that disrupts strong semantic inter-modality correlations, facilitating a deeper understanding of emotional consistency.
\item Our empirical investigations, conducted on the IEMOCAP and MSP-IMPROV datasets under both missing and full modality conditions, affirm the superior performance of our proposed CM-ARR framework.
\end{itemize}

\section{Related Work}

\subsection{Incomplete Multimodal Learning}
In MMER, there have been remarkable advances in research addressing the modality absence problem. These methods fall into two main categories: missing modality generation \cite{cai2018deep,suo2019metric,du2018semi} and multimodal joint representation learning methods \cite{pham2019found,han2019implicit,yuan2021transformer}.

\textbf{Missing Modality Generation} aims to utilize available modalities to predict or reconstruct missing ones. \citet{Tran2017missing} introduce a Cascaded Residual Autoencoder (CRA) to fill in data with missing modalities. This method effectively recovers incomplete data by integrating a series of autoencoders in a cascaded structure and leveraging a residual mechanism to address corrupted data. Similarly, \citet{cai2018deep} develop a 3D encoder-decoder network that captures the intermodal relationships and compensates for missing modalities through adversarial and classification losses.

\textbf{Multimodal Joint Representation Learning} seeks to learn latent representations in a common feature space from available data that remain robust even when some modalities are missing. \citet{pham2019found} introduce a method for learning robust joint representations through cyclic translation between modalities, thereby enhancing the model's capability to comprehend and represent multimodal data. \citet{zhao2021missing} propose the Missing Modality Imagination Network (MMIN), a unified model designed to address the issue of uncertain missing modalities. \citet{zeng2022robust} employ a Tag-Assisted Transformer Encoder (TATE) network, which guides the network to focus on different missing cases by encoding specific tags for the missing modalities. Furthermore, they \cite{zeng2022mitigating} propose an Ensemble-based Missing Modality Reconstruction (EMMR) framework to detect and recover semantic features of the key missing modality. However, these methods do not consider the effect of heterogeneous modal gaps on missing modality reconstruction and emotion recognition. IF-MMIN \cite{zuo2023exploiting} and DiCMoR \cite{wang2023distribution} work on this problem. The former learns modality-invariant features, and the latter transfers distributions from available modalities to missing modalities to maintain distribution consistency in the recovered data. Nevertheless, these approaches only partially bridge modal gaps and overlook semantic uncertainties across modalities. To overcome these limitations, we introduce the CM-ARR framework, which leverages Gaussian distributions to both align modalities and model semantic uncertainty.

\subsection{Contrastive Learning}
Contrastive learning (CL) \cite{khosla2020supervised,he2020momentum} aims to foster efficient data representations by drawing similar samples closer and distancing dissimilar ones. In recent years, CL has become a cornerstone in the field of representation learning \cite{radford2021learning,wang2021pico,ghosh2022mmer,sun2023using}. Notably, \citet{ji2023map} address the heterogeneity between image and text modalities using unsupervised CL. \citet{pan2023gemo} employ supervised contrastive learning to enhance emotional representation learning by clustering similar text and speech modality samples. Building on these principles, our work introduces both unsupervised distribution-based contrastive learning and supervised point-based contrastive learning. These approaches are designed to bridge the gaps between heterogeneous modalities and decipher common emotional patterns for improved prediction accuracy.

\section{CM-ARR}
\begin{figure*}[t]
  \centering
  \includegraphics[width=6.3in]{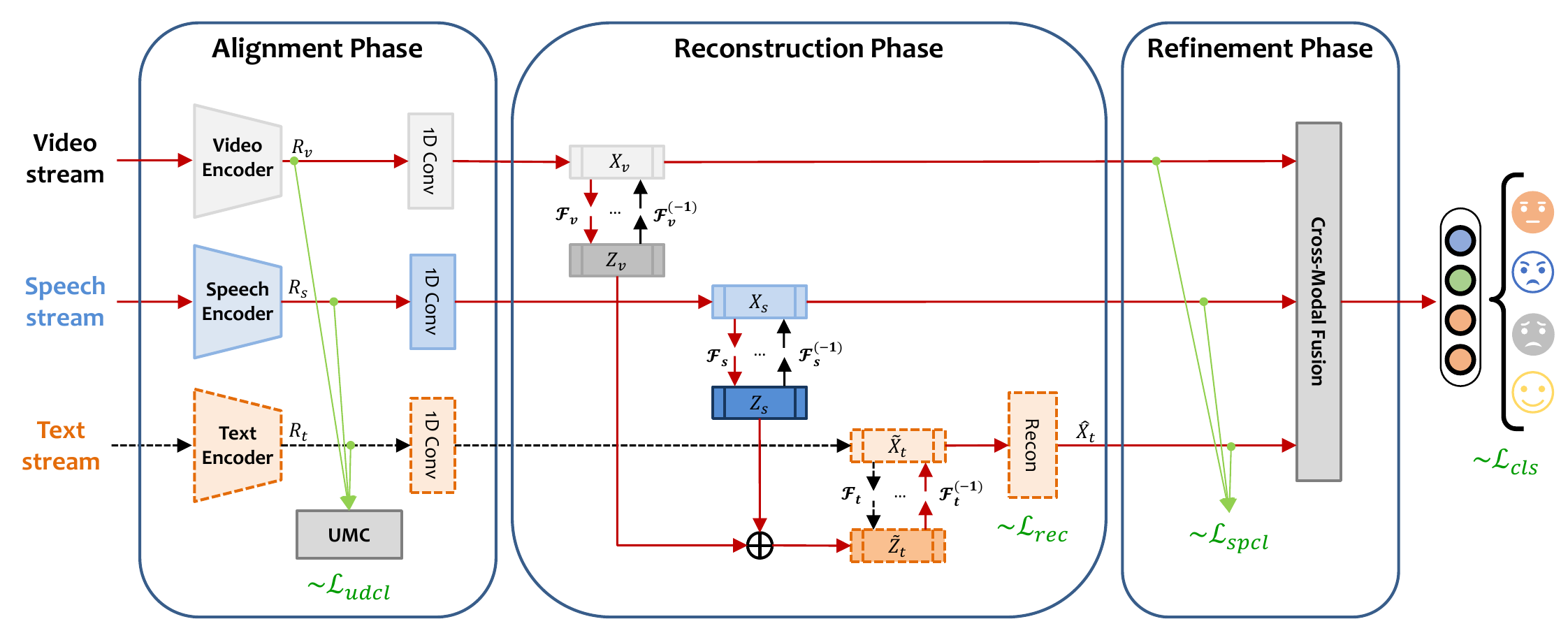}
  \caption{The framework of CM-ARR consists of three phases: the alignment phase employs unsupervised distribution-based contrastive learning to semantically align the video, speech, and text modalities (see UMC in Fig.~\ref{fig:framework3}); the reconstruction phase applies normalizing flow models to each modality; the refinement phase utilizes supervised point-based contrastive learning to accentuate emotional traits. The red arrows denote the inference process assuming the text modality is missing.}
  \label{fig:framework2}
\end{figure*}

In this section, we detail the proposed CM-ARR framework. Fig.~\ref{fig:framework2} illustrates the architecture of CM-ARR, which comprises three main phases: alignment, reconstruction, and refinement. Without loss of generality, we consider a multimodal dataset consisting of three modalities: text, speech,  and video.

\subsection{Alignment Phase}

\subsubsection{Feature Extraction}
Given a speech signal, video segment and its corresponding transcribed text, we extract high-level features for each modality as follows:\\
\textbf{Text Representation:} For each text sequence, we obtain high-level text features $R_t$ using the pre-trained Bert-base model \cite{devlin2018bert}, which has 12 encoder layers, each with 12 self-attention heads and 768 hidden units.\\
\textbf{Speech Representation:} For each speech signal, we obtain high-level speech features $R_s$ using the pre-trained Wav2vec2-base model \cite{baevski2020wav2vec}, where the pre-trained Wav2vec2-base model has 12 encoder layers, each with 8 self-attention heads and 768 hidden units.\\
\textbf{Video Representation:} For each video segment, we utilize a pre-trained DenseNet model \cite{huang2017densely} to extract facial expression features $R_v$, trained on the Facial Expression Recognition Plus (FER+) dataset \cite{barsoum2016training}. These features, referred to as "Denseface," are frame-level sequential features derived from detected faces in video frames, with each feature vector comprising 342 dimensions.

\subsubsection{Unsupervised Distribution-based Contrastive Learning}
In the alignment phase, to mitigate the gaps between heterogeneous modalities while also modeling the uncertainty, we introduce an uncertainty modeling component (UMC) that employs Gaussian distributions to capture semantic uncertainty. This is coupled with unsupervised distribution-based contrastive learning to bring the modal distributions closer.

\begin{figure}[t]
  \centering
  \includegraphics[scale=0.17]{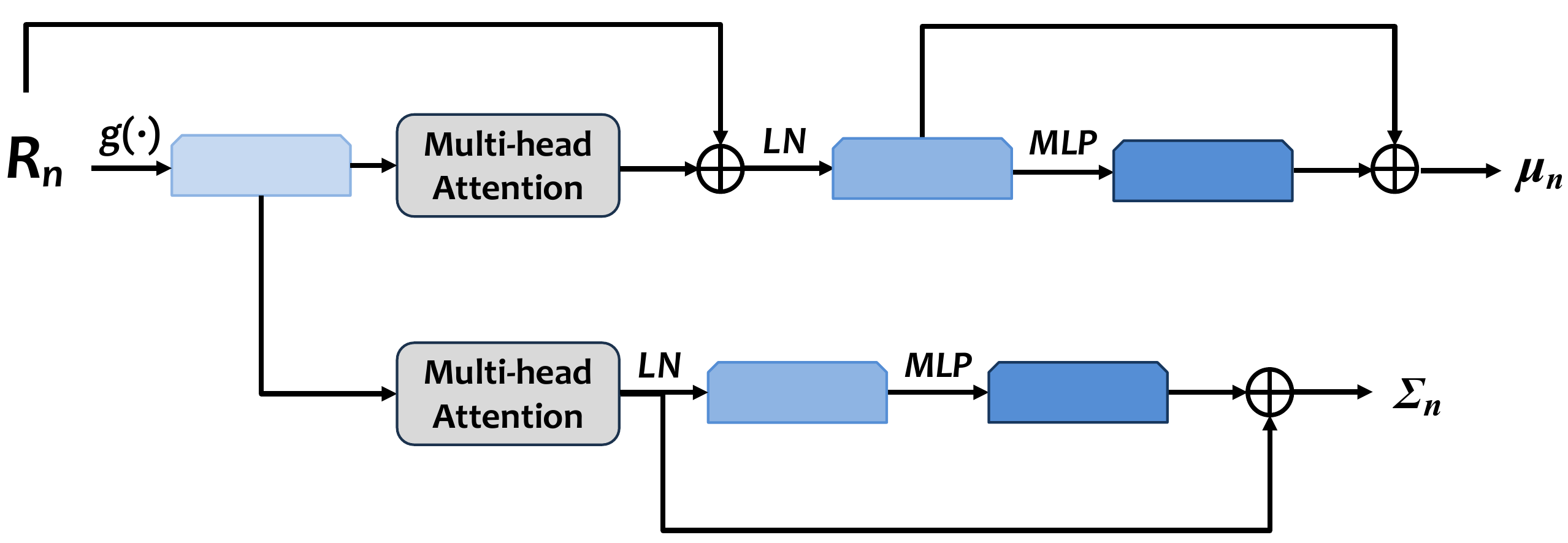}
  \caption{The overall structure of the proposed UMC, where $g(\cdot)$ denotes the gelu function, $LN$ signifies the LayerNorm operation, and $MLP$ indicates the feed forward layer.}
  \label{fig:framework3}
\end{figure}

Fig.~\ref{fig:framework3} illustrates the architecture of the UMC, tasked with learning a Gaussian distribution for each point-based modality representation, $R_n$, where $n \in \{s, v, t\}$. We specifically employ the $MLP$ and the multi-head attention mechanism to enhance feature-level and sequence-level interactions, respectively. The UMC learns a mean vector $\mu_n$ and a variance vector $\Sigma_{n}$ for each $R_n$, transforming the point-based modality representations into Gaussian distributions.

We then implement an unsupervised distribution-based contrastive learning approach to align heterogeneous modal distributions effectively, utilizing the 2-Wasserstein distance \cite{kim2021vilt} to measure the distance between the Gaussian distributions of three modalities:
\begin{eqnarray}
D_{2 W} =\left\|\mu_{n}-\mu_{m}\right\|_{2}^{2}+\left\|\Sigma_{n}-\Sigma_{m}\right\|_{2}^{2},
\end{eqnarray}
where $m, n \in \{s, v, t\}$ and $n \cap m = \varnothing$.

Suppose there are $N$ speech-text, text-video, and speech-video pairs in each batch, where modalities from the same instance are treated as positive samples and those from different instances as negatives. Taking the example of speech-text pairs, we utilize InfoNCE loss \cite{he2020momentum} to compute the loss $\mathcal{L}_{udcl}$:
\begin{align}
&\mathcal{L}_{udcl} =  \mathcal{L}_{nce}^{s 2 t} + \mathcal{L}_{nce}^{t 2 s},\\
&\mathcal{L}_{nce}^{s 2 t} = -\log \frac{\exp \left(S\left(s_i, t_i\right) / \tau\right)}{\sum_{n=1}^{N} \exp \left(S\left(s_i, t_n\right) / \tau\right)},\\
&\mathcal{L}_{nce}^{t 2 s} = -\log \frac{\exp \left(S\left(t_i, s_i\right) / \tau\right)}{\sum_{n=1}^{N} \exp \left(S\left(t_i, s_n\right) / \tau\right)},\\
&S(t, s) =  a \cdot D_{2 W}+b,
\end{align}
where $\tau$ represents a learned temperature parameter. $S\left(\cdot,\cdot\right)$ denotes the similarity between a speech-text pair. $a$ is a negative scale factor and $b$ is a shift value.

\subsection{Reconstruction Phase}
Using representations from different feature spaces directly may lead to inefficacy due to scale, distribution, and semantic inconsistencies. Hence, we initially project each modality's features into a common dimensional space using a 1D convolutional layer, obtaining $X_{s}$, $X_{v}$, and $X_{t}$ for further analysis and processing.

Subsequently, we define the normalizing flow model for each modality as $\mathcal{F}_{n}$, with $n \in \{s, v, t\}$, and $\mathcal{F}^{(-1)}_{n}$ representing its inverse transformation. The features of each modality are fed into their respective normalizing flow model, translating the input features from their original complex distributions to a manageable Gaussian distribution ${Z}_{n}$.
\begin{eqnarray}
{Z}_{n} = \mathcal{F}_{n}\left(X_{n}\right).
\end{eqnarray}

Conversely, the Gaussian distribution ${Z}_{n}$ can be transformed back into the complex distribution of the input features via the inverse transformation $\mathcal{F}^{(-1)}_{n}$, giving $\widetilde{X}_{n}$. Assuming text modality is missing and speech and video modalities are available, we input $X_s$ and $X_{v}$ into respective flow models, $\mathcal{F}_{s}$ and $\mathcal{F}_{v}$, to obtain ${Z}_s$ and ${Z}_{v}$. ${Z}_t$ for the missing text modality can then be computed as:
\begin{eqnarray}
\widetilde{{Z}}_{t} \leftarrow\left({Z}_{s}+{Z}_{v}\right) / 2 \sim \mathcal{N}\left(\mu_{t}, \Sigma_{t}\right), 
\end{eqnarray}
where $\mathcal{N}$ denotes the Gaussian distribution. $\mu_{t}$ and $\Sigma_{t}$ represent the mean and covariance of the text Gaussian distribution, respectively. At this point, $\widetilde{{Z}}_{t}$ represents an estimated Gaussian consistent with missing text. $\widetilde{X}_{t}$ featuring the original text distribution is then generated through the inverse process of the text-specific flow:
\begin{eqnarray}
\widetilde{X}_{t}=\mathcal{F}^{(-1)}_{t}\left(\widetilde{{Z}}_{t}\right).
\end{eqnarray}

Finally, the text-specific reconstruction module is used to recover the text features $\hat{X}_{t}$. The module consists of multiple residual channel attention blocks \cite{wei2022prototype}, where we replace the 2D convolutional layer with 1D. The reconstruction loss is computed as: 
\begin{eqnarray}
\mathcal{L}_{\text {rec }}=\left\|\hat{X}_{t}-X_t\right\|_{F}^{2}.
\end{eqnarray}

Similarly, when speech and video modalities is missing, we follow analogous steps, recovering speech and video features by transferring from available text.

\begin{table*}[t]\footnotesize
  \centering
  \renewcommand\arraystretch{1.6}
\begin{tabular}{m{1.5cm}<{\centering}|m{1.3cm}<{\centering}|m{0.8cm}<{\centering}|m{1.1cm}<{\centering}|m{0.8cm}<{\centering}|m{1cm}<{\centering}|cccc|m{0.7cm}<{\centering}}
\hline
\multirow{2}{*}{Datasets} & \multirow{2}{*}{Languages} & \multirow{2}{*}{Year}& \multirow{2}{*}{Subjects}& \multirow{2}{*}{Type} & \multirow{2}{*}{Access} & \multicolumn{4}{c|}{\makecell{Sample size of emotions used}} & \multirow{2}{*}{Total} \\ \cline{7-10}& & & & & &  \multicolumn{1}{c|}{ang} & \multicolumn{1}{c|}{hap}  & \multicolumn{1}{c|}{neu} & sad &\\ \hline\hline
IEMOCAP & English  & 2008 & 5 male, 5 female & Acted  & Licensed & \multicolumn{1}{c|}{1103} & \multicolumn{1}{c|}{1636} & \multicolumn{1}{c|}{1708} & 1084 & 5531 \\
MSP-IMPROV & English & 2017 & 6 male, 6 female  & Acted  & Licensed & \multicolumn{1}{c|}{460}  & \multicolumn{1}{c|}{999}  & \multicolumn{1}{c|}{1733}   & 627   & 3819  \\
\hline
\end{tabular}
\caption{Statistics of benchmark datasets. Note that the IEMOCAP dataset combines happiness and excited emotions into \emph{``hap''}.}
\label{tab:datasets}
\vspace{-0.15cm}
\end{table*}

\subsection{Refinement Phase}
To further promote the emotional information in the modality representations, we incorporate supervised point-based contrastive learning to refine the modality representations of CM-ARR.

\subsubsection{Supervised Point-based Contrastive Learning}

The learned representations still exhibit strong semantic correlations between various modality pairs. To address this, we utilize supervised point-based contrastive learning. Specifically, we treat modalities from different instances but with the same emotion labels (where different instances may have distinct semantics but share similar emotion characteristics) as positive samples, and those from different labels as negatives. This method transforms one-to-one modality relationships into many-to-many, emotion-centric relationships, thereby enabling the network to learn enhanced emotion representations beyond mere semantics.

\subsubsection{Cross-Modal Fusion}
Finally, after obtaining the recovered text features $\hat{X}_{t}$ and the available speech and video features $X_s$ and $X_{v}$, we fuse them into the multimodal representation $H$ using three cross-modal attention blocks and one self-attention block \cite{vaswani2017attention} for the emotion recognition task. The process is as follows:
\begin{eqnarray}
H_{tv} & = &\text {Cross-Attention}_{tv}\left(Q_{t}, K_{v}, V_{v}\right),\\
H_{ts} & = & \text {Cross-Attention}_{ts}\left(Q_{t}, K_{s}, V_{s}\right),\\
H^{\prime} & = & \text {Concat}\left[H_{tv}, H_{ts}\right],\\
H & = & \text {Self-Attention}\left(H^{\prime}\right),
\end{eqnarray}
where $K_{s}, V_{s} = X_s, K_{v}, V_{v} = X_{v}$ and $Q_{t} = \hat{X}_t$.

\subsection{Optimization Objective}
To train our proposed CM-ARR, four loss functions are required: unsupervised distribution-based contrastive learning loss $\mathcal{L}_{udcl}$, supervised point-based contrastive learning loss $\mathcal{L}_{spcl}$, modality reconstruction loss $\mathcal{L}_{rec}$, and emotion recognition loss $\mathcal{L}_{cls}$. In summary, our training loss is defined as:
\begin{eqnarray}
\mathcal{L} = \alpha \mathcal{L}_{udcl} + \beta \mathcal{L}_{spcl} + \lambda \mathcal{L}_{rec} + \mathcal{L}_{cls} ,
\end{eqnarray}
where $\alpha$, $\beta$, and $\lambda$ represent the trade-off factors.

\renewcommand
\arraystretch{1.3}
\begin{table*}[t]\footnotesize
 \centering
\begin{tabular}{cccccccc}
\hline
Dataset& Avail.& CRA& MMIN& IF-MMIN& CIF-MMIN& DiCMoR$^{\textbf{+}}$& Ours\\ \hline
\multicolumn{1}{l|}{\multirow{8}{*}{IEMOCAP}} & $\{t\}$       & 31.15 / 27.96&67.49 / 68.46&66.58 / 67.51&67.97 / 68.93&68.83 / 70.15& \textbf{69.74 / 70.92} \\
\multicolumn{1}{l|}{}                        & $\{s\}$    & 54.58 / 56.79&54.84 / 56.87&56.06 / 58.38&56.26 / 58.46 &71.02 / 72.30& \textbf{74.15 / 75.38} \\
\multicolumn{1}{l|}{}                         & $\{v\}$     & 53.31 / 51.17 &53.17 / 50.28&53.11 / 51.28&51.40 / 51.39&51.60 / 49.03& \textbf{54.06 / 52.42} \\
\multicolumn{1}{l|}{}                         & $\{v, t\}$   & 31.66 / 28.42&72.98 / 73.60 &72.61 / 73.14 &73.00 / \textbf{74.12} &72.14 / 73.36 & \textbf{73.04 /} 74.09 \\
\multicolumn{1}{l|}{}                         & $\{s, v\}$   & 63.12 / 63.93&64.03 / 64.71 &64.80 / 66.49&66.03 / 67.17&72.28 / 73.49& \textbf{75.51 / 76.38} \\
\multicolumn{1}{l|}{}                         & $\{s, t\}$    & 32.88 / 30.13&74.07 / 75.49 &73.77 / 75.48&74.50 / 75.72&76.89 / 77.83& \textbf{78.95 / 79.70} \\
\multicolumn{1}{l|}{} & Avg.& 44.45 / 43.07&       64.43 / 64.90        &    64.48 / 65.38&64.86 / 65.96&68.79 / 69.36&\textbf{70.90 / 71.48 }\\
\multicolumn{1}{l|}{}                         & $\{s, v, t\}$ & - &77.17 / 77.89&77.97 / 78.58&79.26 / 80.34&78.36 / 79.80 & \textbf{79.86 / 81.06} \\  
 \hline
\hline
Dataset& Avail.& CRA& MMIN& IF-MMIN& CIF-MMIN& DiCMoR$^{\textbf{+}}$ & Ours \\\hline
\multicolumn{1}{l|}{\multirow{8}{*}{\begin{tabular}[c]{@{}c@{}}MSP-\\ IMPROV\end{tabular}}} & $\{t\}$   &46.78 / 28.37 &62.08 / 57.30&61.97 / 58.23&62.42 / 58.71 &62.13 / 59.85 &\textbf{63.57 / 61.49} \\
\multicolumn{1}{l|}{}& $\{s\}$    &37.90 / 38.96 &51.60 / 43.35&50.46 / 40.45&50.66 / 40.37&52.78 / 47.17 &\textbf{55.55 / 50.87} \\
\multicolumn{1}{l|}{}& $\{v\}$    & 59.46 / 42.42&60.09 / 45.75&61.68 / 45.29 &61.10 / 46.15&59.84 / 49.33 &\textbf{62.04 / 51.01}  \\
\multicolumn{1}{l|}{} & $\{v, t\}$   & 54.96 / 38.84 &69.37 / 63.94&67.49 / 63.40 &\textbf{69.90 /} 65.36 &67.30 / 64.21 &68.47\textbf{ / 65.88} \\
\multicolumn{1}{l|}{} &$\{s, v\}$ & 57.85 / 47.70&63.74 \textbf{/ 55.91}&62.42 / 53.14&63.75 / 55.21 &62.70 / 52.65&\textbf{63.84 / }54.25 \\
\multicolumn{1}{l|}{} & $\{s, t\}$ &48.57 / 37.97&64.00 / 60.98 &63.25 / 59.91 &63.78 / 60.80 &67.92 / 65.37 &\textbf{69.49 / 66.85}\\
\multicolumn{1}{l|}{}& Avg. & 52.59 / 39.04& 61.81 / 54.53 & 61.21 / 53.40 & 61.93 / 54.43 & 62.11 / 56.43 & \textbf{ 63.82 / 58.39 } \\
\multicolumn{1}{l|}{} & $\{s, v, t\}$ & - &69.70 / 64.89&69.03 / 63.84&72.02 / 67.12&71.69 / 68.08&\textbf{72.37 / 69.69}\\
 \hline
\end{tabular}
\caption{Performance comparison across testing conditions. The values reported in each cell denote WAR / UAR. ``Avail.'' indicates the available modalities. ``Avg.'' indicates average performance across all conditions. The best results are marked in boldface.}
\label{tab:results}
\end{table*}

\section{Experiments}

\subsection{Datasets}

Our experiments utilize two widely adopted datasets: IEMOCAP \cite{busso2008iemocap} and MSP-IMPROV \cite{busso2016msp}. IEMOCAP, collected by the University of Southern California, is a multi-modal emotion corpus comprising 10,039 utterances from 10 actors who express a range of specific emotions. MSP-IMPROV features 7,798 utterances from six sessions with 12 actors, focusing on the exploration of emotional behaviors during spontaneous dyadic improvisations. Additional details about these datasets are available in Table~\ref{tab:datasets}.

\subsection{Experimental Setup}
We evaluate all comparative methods on IEMOCAP and MSP-IMPROV using 5-fold cross-validation and 6-fold cross-validation, respectively. We employ evaluation metrics such as weighted average recall (WAR) and unweighted average recall (UAR) to assess the performance. In all experiments, parameters are configured as: $\alpha$=1.0, $\beta$=0.1, $\lambda$=10, learning rate=1e-5, batch size=16, epochs=100.

We benchmark CM-ARR against several state-of-the-art (SOTA) frameworks for incomplete multimodal emotion recognition, including CRA \cite{Tran2017missing}, MMIN \cite{zhao2021missing}, IF-MMIN \cite{zuo2023exploiting}, CIF-MMIN \cite{liu2024contrastive}, and DiCMoR$^{\textbf{+}}$ \cite{wang2023distribution}. DiCMoR$^{\textbf{+}}$ denotes our enhanced version of the DiCMoR framework, where we substitute the original speech encoder with a pre-trained Wav2vec2 model. This improves reproduction quality, creating a more robust framework. Using Wav2vec2 also enables a fairer comparison to our method, which similarly utilizes Wav2vec2 for speech encoding.

\subsection{Comparison with SOTA Methods}
Table~\ref{tab:results} presents the performance of CM-ARR against SOTA models in terms of WAR and UAR under full and missing modality testing conditions. Across all evaluation metrics, CM-ARR consistently outperforms the competing models, indicating its superior performance.

Further analysis of CM-ARR's recognition performance on the IEMOCAP dataset under various missing modality conditions reveals notable findings. Specifically, when compared to the best SOTA model, DiCMoR$^{\textbf{+}}$, CM-ARR achieves a relative enhancement of 3.07\% and 3.06\% in WAR and UAR, respectively, on average. Notably, CM-ARR demonstrates exceptional performance when at least one modality is present, particularly with the availability of the speech modality, showing a significant relative improvements of 4.41\% and 4.26\% across WAR and UAR, respectively. Conversely, the performance gains are more modest when only the text modality is present. This discrepancy underscores the speech modality's capacity to encapsulate substantial textual information, facilitating the effective reconstruction of textual modality representations from speech. In contrast, the text modality's limited encapsulation of speech-related information results in less effective speech modality reconstruction. In addition, the performance of our method is also optimized compared to SOTA under full modality testing condition. In conclusion, CM-ARR's ability to leverage available modalities for reconstructing missing modalities significantly mitigates the challenges posed by missing modalities, affirming its effectiveness in addressing the missing modality problem in multimodal emotion recognition.

\renewcommand
\arraystretch{1.3}
\begin{table*}[t]\footnotesize
 \centering
\begin{tabular}{cccccccc}
\hline
Dataset& Avail.& Baseline & w/o $\mathcal{L}_{udcl}$ & w/o $\mathcal{L}_{spcl}$ & w/o attention & w/ Point & Ours\\ \hline
\multicolumn{1}{l|}{\multirow{8}{*}{IEMOCAP}}&$\{t\}$    &66.63 / 67.15 &69.44 / 70.20&69.49 / 70.45&68.71 / 69.75&66.71 / 67.96 & \textbf{69.74 / 70.92 } \\
 \multicolumn{1}{l|}{} &   $\{s\}$    &55.57 / 57.75 &71.90 / 72.67&71.89 / 72.99&71.68 / 72.72 &70.48 / 71.74& \textbf{74.15 / 75.38 } \\
 \multicolumn{1}{l|}{} &   $\{v\}$    &42.91 / 37.54&50.24 / 48.35 &51.14 / 48.56 &50.00 / 47.76&50.35 / 48.99 & \textbf{54.06 / 52.42 } \\
 \multicolumn{1}{l|}{} &   $\{v, t\}$ &68.71 / 68.94&71.59 / 72.90 &72.55 / 73.33 &70.72 / 71.53&71.06 / 72.19 & \textbf{73.04 / 74.09 } \\
 \multicolumn{1}{l|}{} &   $\{s, v\}$ &61.10 / 62.84&72.06 / 72.96 &73.37 / 73.86 &72.10 / 73.47&72.65 / 73.26 & \textbf{75.51 / 76.38} \\
 \multicolumn{1}{l|}{} &   $\{s, t\}$ &75.34 / 76.61 &76.51 / 78.04 &77.80 / 78.83 &76.77 / 77.91&75.53 / 76.67& \textbf{78.95 / 79.70 } \\
\multicolumn{1}{l|}{} &   Avg.       & 61.71 / 61.80 & 68.62 / 69.18 & 69.37 / 69.67 & 68.33 / 68.85 & 67.79 / 68.46 & \textbf{70.90 / 71.48 }\\
 \hline
\end{tabular}
\caption{Emotion recognition results of the ablation study evaluating CM-ARR components on IEMOCAP. The values reported in each cell denote WAR / UAR.}
\label{tab:ablation}
\end{table*}

The right side of Table~\ref{tab:results} shows the performance comparison between CM-ARR and SOTA methods on the MSP-IMPROV corpus. Given the corpus's complexity as a challenging sentiment analysis dataset, the performance of these methods is generally modest. However, experimental results indicate that CM-ARR consistently surpasses SOTA methods across various scenarios, demonstrating its superior effectiveness and robust generalization capabilities.
 
\subsection{Ablation Study}
In Table~\ref{tab:ablation}, ablation experiments are conducted on each component of the CM-ARR framework. To illustrate the limitations of models trained exclusively on full modalities in addressing missing modality scenarios, we establish a full modality baseline model, denoted as 'Baseline', which includes feature extraction and cross-modal fusion components. Results from the IEMOCAP dataset indicate a significant performance decline in the Baseline model when faced with missing modalities, underscoring its vulnerability to conditions of modality absence, given its training on the presumption of modality completeness.

\textbf{Effects of Unsupervised Distribution-based Contrastive Learning:} To verify the effectiveness of the alignment phase, we perform an ablation experiment (w/o $\mathcal{L}_{udcl}$) to evaluate the performance, as shown in Table~\ref{tab:ablation}. The results show that the models with unsupervised distribution-based contrastive learning achieve better performance. This suggests that distribution-based representations could learn richer semantic information from modal uncertainty and help bridge the distributional divergences between modalities, which facilitates subsequent reconstruction. Additionally, replacing distribution-based contrast learning with point-based representation (w/ Point) further demonstrates that leveraging modal uncertainty to gather diverse semantic information offers added advantages. Consequently, Gaussian distribution-based representations prove superior to instance-based representations.

\textbf{Effects of Supervised Point-based Contrastive Learning:} To validate the effectiveness of the refinement phase, we present the results of an ablation experiment (w/o $\mathcal{L}_{spcl}$). The results show that it is helpful to improve the performance by disrupting the semantic correlation between modalities through our point-based supervised contrast learning. This method allows the model to capture more generalized patterns within modal information, reducing the risk of overfitting to specific semantic content. Consequently, this approach emphasizes emotionally significant attributes, thereby enhancing the representation's robustness.

\begin{figure*}[htb]
\centering
\includegraphics[scale=0.50]{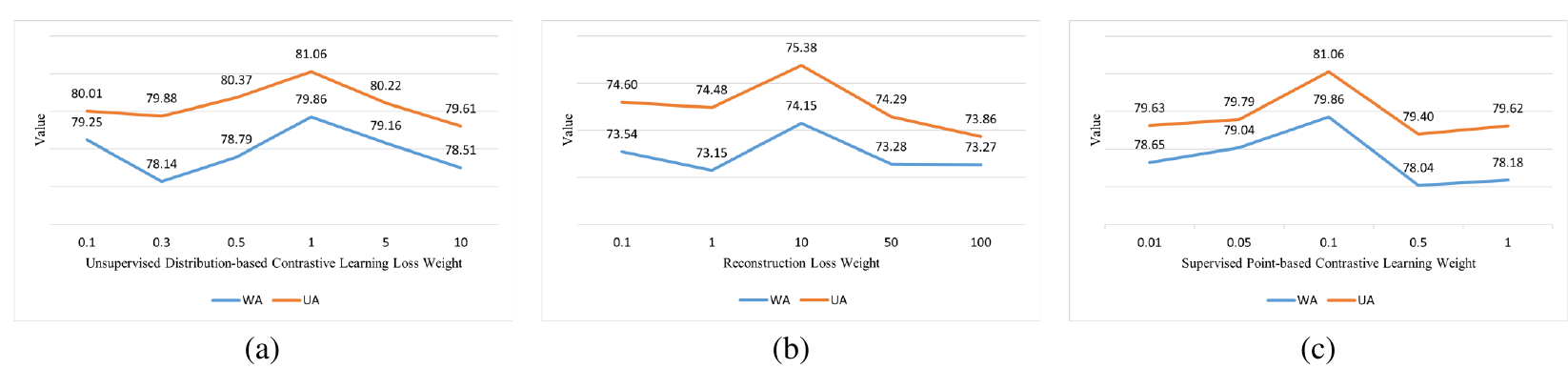}
\caption{The effect of weights $\alpha, \beta$, and $\lambda$ on performance.}
\label{fig:para}
\end{figure*}

\begin{figure*}[htb]
\centering
\includegraphics[scale=0.43]{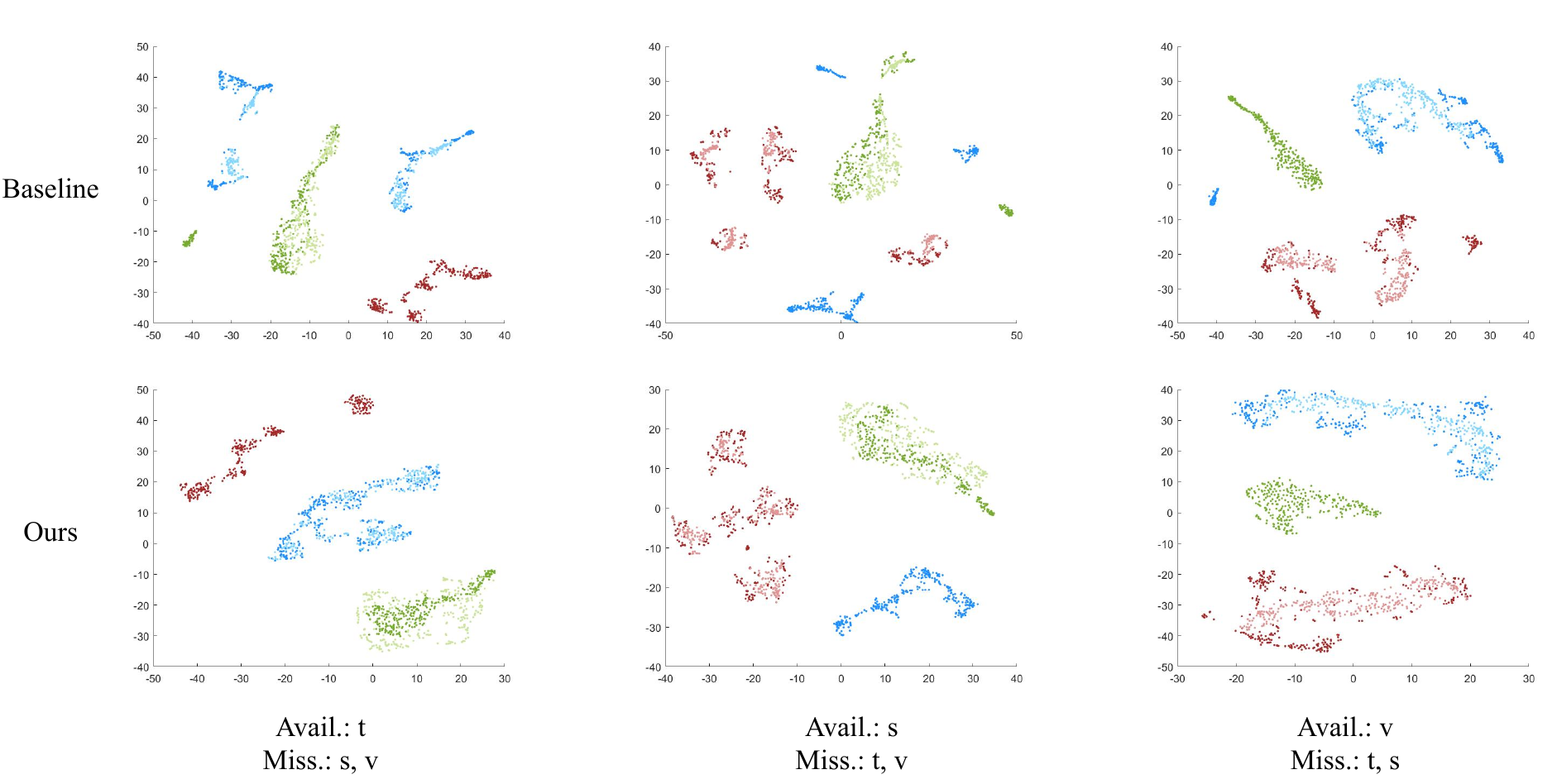}
\caption{Visualization of the representations from different methods on the IEMOCAP corpus test set. Light blue represents speech reconstruction representations, while light red and light green depicts text and video reconstruction representations, respectively, with their corresponding darker shades indicating ground truth. }
\label{fig:flowchart}
\end{figure*}

\subsection{Parameter Analysis}
To thoroughly investigate the impact of various parameter settings on model performance, we conduct the comparative analysis focusing on the effects of $\mathcal{L}_{udcl}$,  $\mathcal{L}_{rec}$, and $\mathcal{L}_{spcl}$ weights.

In Fig.~\ref{fig:para} (a), the UAR (orange line) demonstrates relatively stable performance compared to WAR. This indicates that a moderate loss weight (1.0) yields the best performance for both WAR and UAR in unsupervised distribution-based contrastive learning. Similarly, the impacts of reconstruction loss and supervised point-based contrastive learning weights on performance are illustrated in Fig.~\ref{fig:para} (b) and (c).

In summary, all three loss weights significantly influence model performance. The optimal weights are 1.0 for $\mathcal{L}_{udcl}$, 10.0 for $\mathcal{L}_{rec}$, and 0.5 for $\mathcal{L}_{spcl}$.

\subsection{Visualization Analysis}
In Fig.~\ref{fig:flowchart}, we use t-SNE \cite{van2008visualizing} to visualize the distribution of the modality representations for Baseline and our CM-ARR. 

Fig.~\ref{fig:flowchart} illustrate the impact of CM-ARR on the reconstructed representation in scenarios of modality absence. In baseline, there is noticeably less overlap between the reconstructed modality and its ground-truth representations, with the distribution shape of the reconstructed representations markedly differing from that of the ground-truth representations. In contrast, ours (CM-ARR) demonstrates that the distributional similarity of reconstructed representations to the ground-truth representations is significantly enhanced, particularly evident in the overlap of clusters and the distribution shapes.

\section{Conclusion}
In this paper, we introduce the Cross-Modal Alignment, Reconstruction, and Refinement (CM-ARR) framework, designed to improve multimodal emotion recognition in incomplete data scenarios. CM-ARR effectively models uncertainty within the semantic space using unsupervised distribution-based contrastive learning, reducing the distributional gap. The reconstruction phase utilizes a normalizing flow model to transform aligned distributions, while the refinement phase augments the emotional content of the reconstructed representations. Extensive validation on the widely recognized IEMOCAP and MSP-IMPROV datasets confirms the superior effectiveness of our approach.

\section*{Limitations}
The efficacy of the CM-ARR framework primarily hinges on its ability to leverage available modalities for reconstructing missing ones, thereby mitigating the adverse impacts of modality absence. The experiments with missing modalities show that different modalities contribute to emotion recognition to different degrees. For example, the video modality in the iemocap corpus is weak with a low degree of its contribution to emotion recognition. Therefore, how to deal with the transformation between weak and strong modalities and measure the importance of these modalities is a more interesting issue, which will motivate us to further optimize our proposed CM-ARR.

\section*{Acknowledgements}
This work was supported in part by NSF China (Grant No. 62271270).

\bibliography{custom}

\end{document}